\documentclass[aps]{revtex4}
\usepackage{epsfig,rotating}
\begin{document} 
\author{\bf H. J. de Vega$^{(a)}$ and N. S\'anchez$^{(b)}$} 
\affiliation 
{(a) LPTHE, Universit\'e Pierre et Marie Curie (Paris VI) et Denis Diderot 
(Paris VII), Tour 16, 1er. \'etage, 4, Place Jussieu, 75252 Paris, Cedex 05, 
FRANCE. Laboratoire Associ\'e au CNRS UMR 7589. \\
(b) Observatoire de Paris,  Demirm, 61, Avenue de l'Observatoire,
75014 Paris,  FRANCE. 
Laboratoire Associ\'e au CNRS UA 336, Observatoire de Paris et
\'Ecole Normale Sup\'erieure.  }
\title{\bf THE STATISTICAL MECHANICS OF THE SELF-GRAVITATING GAS:
EQUATION OF STATE AND FRACTAL DIMENSION}
\date{June 2000}
\begin{abstract} 
We provide a complete picture of the self-gravitating non-relativistic
gas at thermal 
equilibrium using Monte Carlo simulations (MC), analytic mean field 
methods (MF) and low density expansions. The system is shown to
possess an infinite volume limit, both in the canonical (CE) and in
the microcanonical ensemble (MCE) when $N, V  \to \infty$, keeping
$N/ V^{1/3}$ fixed. We {\bf compute} the equation of state (we do not assume
it as is customary), the entropy, the free energy, the chemical potential,
the specific heats, the compressibilities,  the speed of sound and
analyze their properties, 
signs and singularities. The MF equation of state obeys a
{\bf first order} non-linear differential equation of Abel type.
The MF gives an accurate picture in agreement with the MC simulations
both in the CE and MCE.  The
inhomogeneous particle distribution in the ground state suggest a
fractal distribution with Haussdorf dimension $D$ with $D$ slowly decreasing
with increasing density, $ 1 \lesssim D < 3$.
\end{abstract} 
\date{June 2000}
\maketitle
The study of the self-gravitating thermal gas has both fundamental and
practical physical interest\cite{ant,nos}: (i) It possess remarkable
thermodynamic properties due to the long range  of the
gravitational force [such properties never occur in ordinary systems
with short range forces]. (ii) It plays a central r\^ole in
astrophysics and cosmology; cold  clouds in the interstellar medium
and its remarkable observed scaling laws (from $10^{-4}$pc up to $100$pc),
as well as galaxy distributions (up to $100-200$Mpc) can be described with it. 
(iii) It is relevant for the study of stellar objects in the non-relativistic
and relativistic cases.

The ground state is {\bf inhomogeneous} and the
usual thermodynamic limit:  number of particles $N \to \infty$, volume
$V \to \infty$, with $N/V$ fixed  leads to  collapse
into a very dense phase. The gaseous phase can only exist when $N, V
\to \infty$ with $N/ V^{1/3}$ fixed. This is a {\bf diluted} limit
where the particle density $N/V$ goes 
to zero as $V^{-2/3}$. The appropriate dimensionless variable (for
particles of  mass $m$ at temperature $T$ on a  box of volume $V$)
is $\eta \equiv {G \, m^2 N \over L \; T}$ (it can be  a
spherical or cubic box with  volume $V=L^3$ or $V=(4\pi/3)R^3$,
respectively). $ \eta $ is related to 
the Jeans' length of the gas $ d_J $ through   $ \eta = 3 (
L/d_J )^2 $. We call `thermodynamic limit' 
the limit  $N \to \infty,  R \to\infty$ with a fixed ratio 
$R/N$. In this limit, physical magnitudes are expressed 
naturally  as functions of  $\eta$. 

For small  $\eta$, the gas behaves as a perfect gas. For growing 
 $\eta ,\,   PV/NT$ (where $P$ is the pressure) decreases
(see figs.1-2) due to  the attractive character of
gravity. Finally, at some critical   $\eta_{crit}$ the gas exhibits
a sharp clumping  transition to a dense phase with {\bf negative
pressure}. The extension of the gaseous phase and the value $\eta_{crit}$
 depend on the thermodynamical ensemble 
 (see figs.1-2): the gas phase is  larger in the
microcanonical ensemble (MCE) and  smaller in the  canonical 
ensemble (CE). We investigate the self-gravitating gas with Monte
 Carlo (MC)  and analytic  mean
field (MF) methods; in the dilute limit we expand in powers of $\eta$.
Our results show that  the  CE and MCE yield the same results in
their common range of the gaseous phase. The MF correctly describes the
thermodynamic limit except near the critical points, 
the MF is valid for $N|\eta-\eta_{crit}|\gg 1$. The vicinity of the
critical point should be studied in a double scaling limit $N \to
\infty,\; \eta \to \eta_{crit}$. 

The particle distribution $\rho({\vec q})$ proves to be {\bf inhomogeneous}
(except for $\eta \ll 1$) and described by an universal function of
the geometry,  $\eta$ and the ratio ${\vec r} = {\vec q} / R$. 
$ D $ slowly decreases from the value $ D = 3 $ for the ideal gas
($\eta=0$) till $ D = 0.98 $ in the extreme limit of the MC point
taking the value $1.6$ at $\eta_{crit}$ [see Table 1]. The particle density in
the bulk behaves as $\rho({\vec q}) \simeq r^{D-3}$. 
This indicates the presence of a fractal distribution with
Haussdorf dimension $D$. 

In the MCE, the entropy can be written as
\begin{eqnarray}\label{smc}
e^{S(E,N)_{MCE}}  &=& {1 \over N !}\int\ldots \int
\prod_{l=1}^N{{d^3p_l\, d^3q_l}\over{(2\pi)^3}}\;
\delta\!\left[E - \sum_{l=1}^N\;{{p_l^2}\over{2m}} - 
U({\vec q}_1, \ldots  {\vec q}_N) \right] 
\end{eqnarray}
where $E$ is the total energy, $G$ is Newton's gravitational
constant and
$$
U({\vec q}_1, \ldots {\vec q}_N) \equiv  - G \, m^2 \sum_{1\leq l < j\leq N} 
{1 \over { |{\vec q}_l - {\vec q}_j|}} \; .
$$
In the CE the  partition function  can be written as 
\begin{equation}\label{fp}
{\cal Z}_{CE}\!\! =\!\! {1 \over N !}\!\!\int\!\!\ldots\!\! \int
\prod_{l=1}^N\;{{d^3p_l\, d^3q_l}\over{(2\pi)^3}}\; e^{- \beta \left
[\sum_{l=1}^N {{p_l^2}\over{2m}} + U({\vec q}_1, \ldots {\vec
q}_n)\right]}  \, .
\end{equation}
We make now explicit the volume dependence by introducing the 
variables ${\vec r}_l ,\;  1\leq l \leq N$ as
${\vec q}_l = R \; {\vec r}_l \; , \; {\vec r}_l =(x_l,y_l,z_l),$
where $|{\vec r}_l| \leq 1$ for a spherical box and $0\leq
x_l,y_l,z_l \leq 1$ for a cubic box. The momentum integrals can be
straightforwardly computed both in eq.(\ref{smc}) and in eq.(\ref{fp}) yielding
\begin{eqnarray}\label{sN}
&&e^{S(E,N)_{MCE}}  = {(m\,N)^{3N-2}  \, R^{3N/2 +1} \, G^{3N/2 -1} 
\over N !\, \Gamma\left( {3N \over 2} \right)\, {2\pi}^{3N/2}} \;
\int \ldots \int \!\! \prod_{l=1}^N d^3x_l \; 
\left[\xi + {1 \over N}u({\vec r}_1,\ldots,{\vec r}_N) \right]^{3N/2
-1}_+ \; , \cr \cr
&&{\cal Z}_{CE} = {1 \over N !}\!\left({m T R^2\over{2\pi}}\right)^{\frac{3N}2}
\! \int \!\!\!\ldots\! \int\!\! 
\prod_{l=1}^N d^3r_l\;\; e^{ \eta \; u({\vec r}_1,\ldots,{\vec
r}_N)}\label{fp2} 
\end{eqnarray}
where $u({\vec r}_1,\ldots,{\vec r}_N)   \equiv \frac1{N} \sum_{1\leq
l < j\leq N} {1 \over { |{\vec r}_l - {\vec r}_j|}}$ 
and $\xi \equiv { E \, R \over G \, m^2 \, N^2}$
are dimensionless and we introduced the notation $[ X ]^n_+ \equiv
X^n\; \theta(X)$,  being $\theta(X)$ the step function. All space
integrals in eq.(\ref{fp2}) and below are over an unit volume.
The temperature is given in the MCE  as a function of $E$ and $\xi$
by (for $N \gg 1$)
\begin{eqnarray}\label{termo}
&&{1 \over T} = \left( {\partial S \over \partial E} \right)_{\! V} \!\!= {N
\over E} \; g(\xi) \; , \;
 {P V\over NT} = {V \over N} \! \left( {\partial S \over \partial V}
\right)_{\! E}\!\! = \frac12 + \frac13 \; g(\xi)  \\ \cr
&&\qquad g(\xi)\equiv { \xi \over N} {\partial \over \partial \xi}\log  
w(\xi,N) \; , \;
w(\xi,N)\equiv \int \!\!\ldots \int \prod_{l=1}^N\; d^3r_l \; 
\left[\xi + {1 \over N}u({\vec r}_1,\ldots,{\vec r}_N) \right]^{3N/2
-1}_+\; . \nonumber
\end{eqnarray}
In the CE, the free energy results
\begin{eqnarray}\label{flib}
&&F = -T \log {\cal Z}_{CE} = F_0 - T  \; \Phi_N(\eta) \;, \cr \cr
&&\Phi_N(\eta) = \log \left[
\int \!\!\ldots \int\!\!
\prod_{l=1}^N d^3r_l\;\; e^{ \eta \; u({\vec r}_1,\ldots,{\vec r}_N)}
\right]\; ,
\end{eqnarray}
where $F_0 = -N T \log{ e V \over N} \left({mT\over
2\pi}\right)^{3/2}$ is the free energy for an ideal gas. Defining
$f(\eta) \equiv 1 - {\eta \, T \over 3 \, V} \; \Phi_N'(\eta)$, we
then find
\begin{eqnarray}\label{pres}
&&P = -\left({ \partial F \over  \partial V}\right)_T = {N
T \over V} \;  f(\eta) \; , \; 
F = F_0- 3NT\!\int_0^{\eta}\! dx\, { 1 - f(x) \over x}\; ,\cr \cr
&&{E \over N T}= - {T \over N} { \partial 
\over  \partial T}\left({F \over T}\right)_V= 3 \; \left[  f(\eta)
-\frac12\right] \; ,\cr \cr
&&{S \over N} = -\frac32 -{ F_0\over NT} + 3   f(\eta) + 3 
\! \int_0^{\eta} \! dx\, { 1 - f(x) \over x} \; . \hskip -2cm
\end{eqnarray}
$P$ stands for the external pressure on the gas. 
Notice that the {\bf equation of state} $PV = NT f(\eta)$
is here {\bf computed} from the gravitational interaction, we do not
assume an equation of state as it is customary. 
It follows from eqs.(\ref{termo}) and (\ref{pres}) that $\eta\, \xi = g(\xi) =
3\,[f(\eta) - 1/2]$ and that the virial theorem, $(3PV = \frac32 NT +
E )$,  holds both in the MCE and in the CE. 

We find that the following properties hold in the $N \to \infty \; ,
L \to  \infty$ limit provided $\eta$ and $\xi$ are kept fixed:

(a) $f(\eta)$ and $g(\xi)$ reach finite limits. [In fact, fixed
$\eta$  implies  a finite limit for $\xi$ and viceversa]. 

(b) The energy, free energy, entropy and $P V$ are proportional to $N$
(besides a $\log N$ term in the entropy as for an ideal gas).   

(c) {\bf Both} ensembles,  the MCE and CE yield  {\bf the
same} physical magnitudes.

We have verified properties (a)-(c) in three ways. First, by direct
calculation in the dilute regime ($ \xi \gg 1 \, , \, \eta\ll 1 $) by
expanding  eqs.(\ref{termo}) and eqs.(\ref{flib}) in powers
of $1/\xi$ and $\eta$ for the MCE and CE, respectively. Second, by
performing Monte Carlo simulations both in the MCE and the CE. Third, by 
mean field  approximations to eqs.(\ref{termo}) and eqs.(\ref{flib}). 

The specific heat at constant volume takes the form,
\begin{equation}\label{ceV}
c_V = {T \over N}  \left({ \partial S \over  \partial T}\right)_V
= 3 \left[  f(\eta)-\eta \; f'(\eta) -\frac12 \right]\; .
\end{equation}
This expression is valid in the thermodynamic limit both in the CE and
MCE. In the CE, the specific heat is related to  the fluctuations of
the potential energy $(\Delta U)^2$ and it is positive defined,
$$
(c_V)_{CE} =\frac32 + (\Delta U)^2 \; , \;
(\Delta U)^2 \equiv {{<U^2>-<U>^2}\over N \; T^2}
$$
\begin{center}
 {\it {\bf Dilute Limit Calculations.}} 
\end{center}
For a dilute gas, we expand the integrands in 
eqs.(\ref{termo}) and eqs.(\ref{flib}) in powers of $1/\xi$ or $\eta$,
respectively. In the $N \to \infty$ limit  the expressions simplify
considerably. Moreover, divergent integrals  in the zero cutoff limit
are eliminated by $1/N$ factors. That is, after the thermodynamic
limit is taken the cutoff corrections are of the order $ {\cal
O}(a^2) $ and clearly vanish in the $ a \to 0 $ limit. We find after
calculation, 
\begin{eqnarray}\label{fgper}
{E \over N T}&=& g(\xi) = \frac32 -{ 9\, b_0\over 2 \xi} - {9 \over
4\, \xi^2}\left(b_1  -  
42 \, b_0^2 \right) + {\cal O}(\xi^{-3})\cr \cr
{PV \over N T}&=&f(\eta) = 1 -  b_0 \; \eta -\eta^2\left[ \frac13 b_1
- 12 \,  b_0^2 \right] +  {\cal O}(\eta^3)\; .
\end{eqnarray}
where
$$
b_0 = \frac16 \int \!\! \int \!\! { {d^3 r_1 \; d^3 r_2} \over
{ |{\vec r}_1 - {\vec r}_2|}}\; , \;
b_1 = \int\!\! \int \!\! { {d^3 r_1 \; d^3 r_2\; d^3 r_3} \over
{ |{\vec r}_1 - {\vec r}_2| |{\vec r}_1 - {\vec r}_3|}}
$$
For a sphere of unit volume we find $b_0^{sph} = 1/5 
\, ,  b_1^{sph} = 153/35$, whereas for a unit volume cube $b_0^{cube} =
0.19462 \ldots  .$ Notice that the cube and sphere values differ only
by about $3\%$. We see from eq.(\ref{fgper}) that the perfect gas
behaviour gets corrected by {\bf negative}  definite terms due
to the attractive nature of the gravitational interaction. 
[The full correction to the ideal gas, $\Phi_N'(\eta)$, is positive definite
according to eq.(\ref{flib})].

\begin{center}
{\bf Monte Carlo Calculations.}
\end{center}
We  applied  the standard Metropolis algorithm in a cube of size $L$
with total energy $E$ in the MCE and at temperature $T$ in the CE. The
number of particles $N$ went  up to $2000$\cite{montec}. We introduced
a small short distance cutoff $ \sim 10^{-3} L - 10^{-6} L$ in the
attractive Newton's potential. All results in the gaseous phase were
insensitive to the cutoff value. 

Two different phases show up: a non-perfect gas for $\eta < \eta_c$,
and a condensed system with {\bf negative} pressure for $\eta >
\eta_c$. The transition between  the two phases 
is very sharp, with a negative jump in the entropy from the gas to the
condensed phase. This phase transition is associated with the Jeans
instability.

We plot in figs.1-2 $f(\eta) = PV/[NT]$ as a function of $\eta$. 
For small $\eta$,  the MC results for  $PV/[NT]$
 reproduce very well the analytical formula (\ref{fgper}).  $PV/[NT]$
monotonically  
decreases with  $\eta$ as forecasted by the dilute expansion
(\ref{fgper}). For $\eta = \eta_T \simeq 1.51$  (point $T$ in fig.2),
close to $\eta_C \simeq 1.54$ (point $C$ in fig.2) 
a phase transition suddenly happens in the CE and  $PV/[NT]$ becomes
large and negative. The  interparticle distance $<r>$  monotonically decreases
with  $\eta$ too. When   $\eta$ crosses $\eta_T \; , <r>$ has a sharp
decrease.  

The MCE and CE Monte Carlo results are very close (up to the
statistical error) for $0 < \eta <\eta_C$, that is for $\infty >
\xi > \xi_C \simeq -0.19$. In the MCE the gas does not clump at $\eta
= \eta_C$ (point $C$ in fig.2) and the specific heat becomes negative
between the points $C$ and $MC$. The gas does clump in the MCE at
$\xi \simeq -0.32 \; , \; \eta \simeq 1.35 $ (point $MC$ in fig.2)
increasing both its temperature and pressure discontinuously. As is
clear, the domain between $C$ and $MC$ cannot be reached in the CE
since $c_V > 0$ in the CE.

\begin{center}
{\bf Mean Field Calculations.}
\end{center}
We now recast the coordinate partition function $e^{\Phi_N(\eta)}$
as a functional integral in the thermodynamic limit. 
\begin{eqnarray}\label{zcanmf}
&&e^{\Phi_N(\eta)} \buildrel{ N \gg 1}\over= \int\int D\rho\; da \;
e^{-N s[\rho(.)]+ia\left(\int  d^3x \, \rho({\vec x}) - 1 \right)}
\\ \cr
&&\!\!s[\rho(.)]= -{\eta \over2}\!\int \!\! {d^3x d^3y \over |{\vec
x}-{\vec y}|} 
\rho({\vec x}) \rho({\vec y}) +\! \int \!\! d^3x\rho({\vec
x})\log[\rho({\vec x})/e] \;.\nonumber 
\end{eqnarray}
where we used the coordinates $\vec x$ in the unit volume.
The first term  is the potential energy, the second term is the functional
integration measure for this case (see \cite{lipa}). 
Eq.(\ref{zcanmf}) is dominated for large $N$ by
its stationary point solutions:
\begin{equation}\label{eqrho}
\log \rho({\vec x}) - \eta \int{ d^3y \; \rho({\vec y})\over |{\vec
x}-{\vec y}|} =a\; ,
\end{equation}
$a$ is a Lagrange multiplier enforcing 
the constraint $\int  d^3x \, \rho({\vec x}) = 1$.
Applying the Laplacian and setting $\phi({\vec x}) \equiv
\log\rho({\vec x})$ yields, for  the spherically symmetric case
\begin{equation}\label{eqexpr}
{d^2\phi\over dr^2} + \frac2{r} {d \phi\over dr} +  4\pi \eta^R \;
e^{\phi(r)} = 0\; .
\end{equation}
with  boundary conditions $\phi'(0)=0$ and  $\phi'(1)= -\eta^R $. 
Here  the variable $ \eta^R $ appropriate for a spherical
symmetry is defined as $ \eta^R \equiv {G \, m^2 N \over R \; T}
= \eta \; \left({4\pi\over 3}\right)^{1/3} =  1.61199\ldots \; \eta\; $ .

Using the scale covariance of eq.(\ref{eqexpr}) \cite{nos},
$\phi(r)$ can be expressed as $\phi(r) = \log(\lambda^2/ 4\, \pi \,\eta^R) +
\chi(\lambda \, r)$ where 
\begin{equation}\label{ecuaxi}
\chi''(\lambda) + {2 \over \lambda} \, \chi'(\lambda) +
e^{\chi(\lambda)} = 0 ,  \; \chi'(0) = 0 
\end{equation}
This equation is invariant under the transformation:
\begin{equation}\label{invchi}
\lambda \Rightarrow \lambda \; e^{\alpha} \quad \, \quad 
\chi(\lambda)\Rightarrow \chi(\lambda) - 2 \; \alpha \; ,
\end{equation}
where $ \alpha $ is a real number. Hence, we can set 
$ \chi(0) \equiv 0 $ without loosing generality.

$\chi(x)$ is independent of
$\eta^R$ and  $\lambda$ is related to $\eta^R$ by $\lambda \;
\chi'(\lambda ) = -\eta^R$.

Evaluating the functional integral in eq.(\ref{zcanmf}) by saddle point yields
\begin{equation}\label{silla}
e^{\Phi_N(\eta^R)} \buildrel{ N \gg 1}\over= {e^{-N\,s(\eta^R)} \over
\sqrt{D_C(\eta^R)}} \left[ 1 + {C(\eta^R)\over N} + {\cal O}({1 \over N^2})\right]
\end{equation}
where $D_C(\eta^R)$ stands for the determinant of small fluctuations
around the spherically symmetric saddle point (\ref{eqexpr}) and
$C(\eta^R)$  for the two-loop corrections. $D_C(\eta^R)$ can be expressed
as an infinite product over the partial waves $D_C(\eta^R) = \prod_{l\geq
0} [D_l(\eta^R)]^{2l+1}. \, D_0(\eta^R)$ and  $D_1(\eta^R)$ can be
computed in closed form \cite{prox} 
\begin{eqnarray}
 D_0(\eta^R) &=& \frac1{2\eta^R}[ \lambda^2(\eta^R) \,
 e^{\chi(\lambda(\eta^R))}\,  - \,  \eta^R] \; , \; D_1(\eta^R) =
 e^{\chi(\lambda(\eta^R))} \; .\nonumber
\end{eqnarray}
$D_0(\eta^R)$ [see fig.1] is positive for 
$0 \leq \eta^R \leq \eta^R_C = 2.51755\ldots$ and vanish linearly in
$\sqrt{ \eta^R_C-\eta^R}$ at $\eta^R=\eta^R_C$.

We thus get for the free energy in the CE from
eqs.(\ref{flib}),  (\ref{zcanmf}),  (\ref{eqrho}) and (\ref{silla})
\begin{eqnarray}\label{fcampm}
&&F = F_0 +  NT\, s(\eta^R) + \frac{T}2\log D_C(\eta^R)+ {\cal O}(N^{-1})
\cr &&{PV \over NT} = f_{MF}(\eta^R)+ {\eta^R \; D_C'(\eta^R)\over 6 N
D_C(\eta^R)}+ {\cal O}(N^{-1}) \\ \label{f2}
&&f_{MF}(\eta^R) \equiv  { \lambda^2(\eta^R) \over 3 \eta^R }\,
e^{\chi(\lambda(\eta^R))}\\ 
&& s(\eta^R) =  3[1 - f_{MF}(\eta^R)]- \eta^R+ \log[3 f_{MF}(\eta^R)/4\pi] \nonumber 
\end{eqnarray}
It follows from eq.(\ref{ecuaxi}) and (\ref{f2}) that $f_{MF}(\eta^R)$
obeys the {\bf first} order equation 
$$
\eta^R(3f-1)f'(\eta^R)+(3f-3+\eta^R) f = 0 \;.
$$
which reduces to an Abel equation of first kind\cite{kam}. 

By expanding the solution of eq.(\ref{ecuaxi})
in powers of $\lambda$ one checks that eq.(\ref{fcampm})  reproduces
eq.(\ref{fgper})\cite{prox}. We plot in fig.1  $f_{MF}(\eta^R)$ as a
function of $\eta^R$ obtained by solving  eqs.(\ref{ecuaxi}-\ref{fcampm}) 
by  the Runge-Kutta method. We find that MC results (both in the MCE
and CE) and the MF results are in {\it excellent agreement}. (This happens
although the  geometry for the MC calculation is cubic while it is
spherical for the MF). 

The clumping phase transition takes place when $D_C(\eta^R)$ vanishes at
$\eta^R = \eta^R_C$. At such point the expansion in $1/N$ breaks down
since the correction  terms become large in eqs.(\ref{silla}-\ref{fcampm}). 
MF applies  when $N|\eta^R_C-\eta^R| \gg  1$. Since,  
$$
{\eta^R \; D_C'(\eta^R)\over 6  D_C(\eta^R)}\buildrel{ \eta^R \uparrow
\eta^R_C}\over= -{\eta^R_C \over 3(\eta^R_C-\eta^R)} \to -\infty \; ,
$$
eq.(\ref{fcampm}) correctly suggests that $PV/[NT]$ becomes {\bf
large and negative} for $\eta^R \uparrow \eta^R_C$ in agreement with
the MC results.  

From eqs.(\ref{eqexpr}-\ref{fcampm}) we obtain the following behaviour
near the point $C$ in fig.1
\begin{eqnarray}\label{mfcri}
f_{MF}(\eta^R) &\buildrel{ \eta^R \uparrow \eta^R_C}\over=& 1/3 +
0.213738\ldots \sqrt{\eta^R_C-\eta^R} + {\cal O}(\eta^R_C-\eta^R) \label{cVc}
\cr
(c_V)_{MF}&\buildrel{ \eta^R \uparrow \eta^R_C}\over=& 0.80714\ldots
(\eta^R_C-\eta^R)^{-1/2} - 0.19924\ldots+ {\cal O}(\sqrt{\eta^R_C-\eta^R})
\end{eqnarray}

\begin{figure}
\begin{turn}{-90}
\epsfig{file=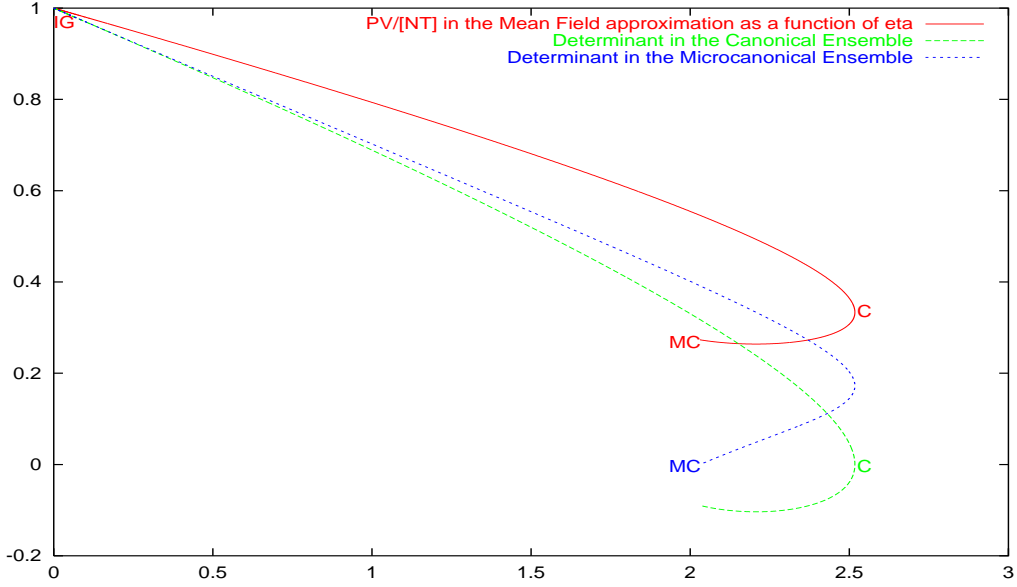,width=8cm,height=14cm} 
\end{turn}
\caption{ $f_{MF}(\eta^R) = P V/[ N T]$ as a function of $\eta^R$ in the MF
approximation [eq.(\ref{fcampm})], $D_0(\eta^R)$ for the canonical
ensemble (CE) and $D^S_{MC}$ for the microcanonical ensemble
(MCE). \label{fig1}} \end{figure}

As noticed before, the CE  only describes  the region between the
points $IG$ (ideal gas point, $\eta = 0$) and $C$ in fig.1. The MCE
goes beyond 
the point $C$ (till the point $MC$) with the physical 
magnitudes described by the second sheet of the square root in
eqs.(\ref{mfcri}). We  have between $C$ and $MC$
\begin{eqnarray}
&& f_{MF}(\eta^R) \buildrel{ \eta^R \uparrow \eta^R_C}\over= 1/3 -
0.213738\ldots 
\sqrt{\eta^R_C-\eta^R} + {\cal O}(\eta^R_C-\eta^R) \cr 
&&(c_V)_{MF}\buildrel{ \eta^R \uparrow \eta^R_C}\over=
-0.80714\ldots(\eta^R_C-\eta^R)^{-1/2} - 0.19924\ldots+{\cal
O}(\sqrt{\eta^R_C-\eta^R})\nonumber 
\end{eqnarray}
$c_V$ is negative in all the interval from $C$ to $MC$ and vanishes at the
point $MC$.

A functional integral representation for $w(\xi,N)$ in the MCE
follows by inverse Laplace transform of $e^{\Phi_N(\eta^R)}$
\cite{prox}. The saddle point in the MCE is $N s(\eta^R)$ as in in the
CE but the  corrections are different. The S-wave determinant results,
$D^S_{MC} = 6 f_{MF}(\eta^R)^2 - (11/2 -\eta^R)f_{MF}(\eta^R) + 1/2$.
It is positive from  $IG$ to $MC$ where it vanishes (fig.1). 

\begin{center}
{\bf Speed of sound and compressibility.}
\end{center}
The isothermal compressibility and the  speed of sound for long
wavelengths\cite{llmf} follow from the equation of state (\ref{pres})
$$
K_T = - { 1 \over V} \left({ \partial V \over  \partial P}\right)_T =
{V \over N\, T} {1 \over {f(\eta^R)+\frac13 \eta^R f'(\eta^R)}} \; ,
$$
\begin{equation}\label{vson}
{v_s^2 \over T}(\eta^R) = { \left[f(\eta^R)-\eta^R f'(\eta^R)\right]^2\over 3
\left[f(\eta^R)-\eta^R f'(\eta^R)-\frac12 \right]} + f(\eta^R)+\frac13 \eta^R
f'(\eta^R)\; .
\end{equation}
$K_T$ is positive for $0 \leq \eta^R < \eta^R_0 = 2.4345...$ where
$K_T$ diverges, it is  negative for $\eta^R_0 < \eta^R < \eta^R_C$. The
singularity of  $K_T$ before but near the  point $C$ appears 
as a preliminary signal of the phase transition and perhaps $\eta^R_0$
is the transition point $T$ seen with the Monte Carlo simulations (see
fig. 2). $K_T$ becomes positive between $C$ and $MC$.   

The speed of sound squared $v_s^2/T (\eta^R)$, is positive and
decreasing in the whole interval between $I$ and $C$. At the 
point $C$ it takes the value $v_s^2 / T(\eta^R_C) = 11/18$. Then,
$v_s^2 / T(\eta^R)$ decreases between $C$ and $MC$ becoming 
negative at $\eta^R_1 = 2.14674...$ $v_s^2 < 0$ indicates an instability
announcing the  $MC$ critical point at $\eta^R_{mc} = 2.03085...$

\begin{center}
{\bf Particle Distribution.}
\end{center}
In the dilute regime $\eta^R \ll 1$ the gas 
density is uniform, as expected. We find that these mass distributions
approximately follow the power law 
\begin{equation}\label{esca}
{ \cal M }(r) \simeq C \; r^D 
\end{equation}
where $ D $ slowly decreases with  $ \lambda(\eta^R) $ as depicted in
Table 2 from the value $ D = 3 $ for the ideal gas ($\eta^R=0$) till $
D = 0.98 $ in the extreme limit of the MC point.

The gravitational potential at the point $ \vec r $ in the MF follows
from eq.(\ref{eqrho}) to be 
$$
U(\vec r) = - { T \over m} \left[ \phi({\vec x}) - a \right] \; .
$$
The {\bf local} pressure in the gas takes then the form $ p(\vec r) = T \,
\rho(\vec r) $. We have thus {\bf derived} the equation of state for the
self-gravitating gas. It turns to be  {\bf locally} the {\bf ideal gas
} equation. But, the self-gravitating gas is inhomogeneous,  the
pressure at the surface of a given volume is not equal
to the temperature times the average density of particles in the volume.
In particular, for the whole volume: $ PV/[NT] = f(\eta^R) \neq 1 $ (except for $
\eta = 0 $). 

\begin{figure}
\begin{turn}{-90}
\epsfig{file=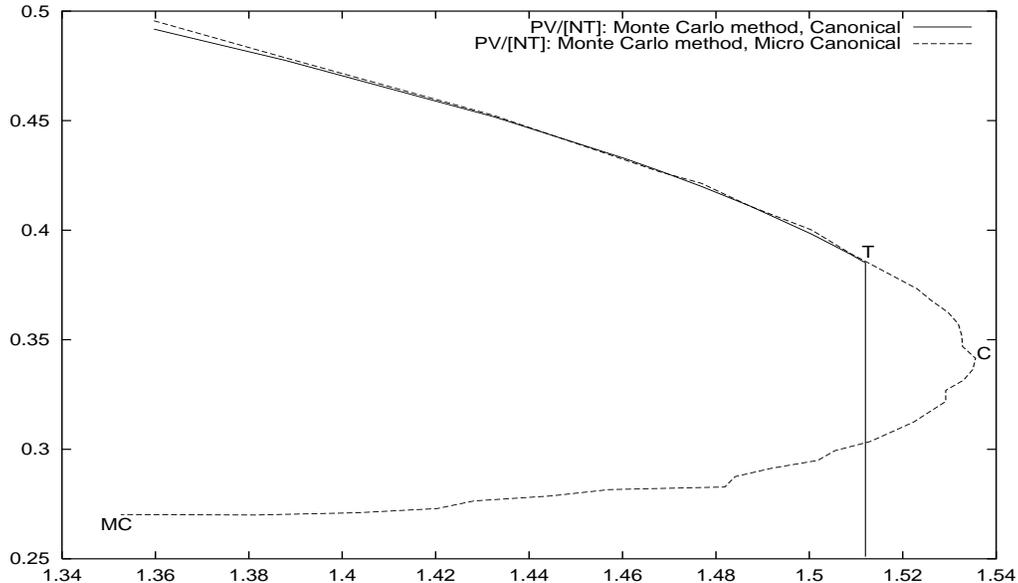,width=8cm,height=14cm} 
\end{turn}
\caption{ $f(\eta) = P V/[ N T]$ as a function of $\eta$  by Monte
Carlo for the MCE and CE ($N=2000$). 
\label{fig2}}
\end{figure}
\begin{tabular}{|l|l|l|}\hline
$ \eta^R $ & $\hspace{0.5cm} D $ & $\hspace{0.5cm} C $\\ \hline 
$ 0.1 $ &    \hspace{0.3cm} $ 2.97 $ \hspace{0.3cm}  &  \hspace{0.3cm}
$ 1.0 $ \hspace{0.3cm}  \\ \hline 
$ \eta^R_{GC} $ & \hspace{0.3cm} $ 2.75  $\hspace{0.3cm} &
\hspace{0.3cm} $ 1.03 $ \hspace{0.3cm} \\ \hline 
 $ 2.0 $  & \hspace{0.3cm} $ 2.22 $ \hspace{0.3cm} &\hspace{0.3cm} $ 1.1
$\hspace{0.3cm} \\ \hline 
$ \eta^R_{C} $ & \hspace{0.3cm} $ 1.60  $\hspace{0.3cm} & \hspace{0.3cm}
$ 1.07 \hspace{0.3cm} $ \\ \hline
$ \eta^R_{MC} $ & \hspace{0.3cm} $ 0.98  $ \hspace{0.3cm}&\hspace{0.3cm}
$ 1.11 $\hspace{0.3cm} \\ \hline 
\end{tabular}

{TABLE 1. The Fractal Dimension $ D $ and the proportionality
coefficient $ C $ as a function of $ \eta^R $ from a fit to the mean
field results according to $ { \cal M }(r) \simeq C \; r^D $.}

We have thus shown that the thermodynamics of the self-gravitating gas
in the $ N \to \infty , \; L \to \infty $ limit with $N/L$ fixed can
be expressed in terms of a single function $ f(\eta)$. Besides
computing $ f(\eta)$ numerically we obtained its Riemann sheet
structure and its behaviour near the critical point analytically
[eq.(\ref{mfcri})].

\end{document}